# Analytic adjoint solution for incompressible potential flows

**Carlos Lozano*** **and Jorge Ponsin**

*Theoretical and Computational Aerodynamics*

*National Institute of Aerospace Technology (INTA), Spain*

**Corresponding author: lozanorc@inta.es*

## Abstract

We obtain the analytic adjoint solution for two-dimensional (2D) incompressible potential flow for a cost function measuring aerodynamic force using the connection of the adjoint approach to Green's functions and also by establishing and exploiting its relation to the adjoint incompressible Euler equations. By comparison with the analytic solution, it is shown that the naïve approach based on solving Laplace's equation for the adjoint variables can be ill-defined. The analysis of the boundary behavior of the analytic solution is used to discuss the proper formulation of the adjoint problem as well as the mechanism for incorporating the Kutta condition in the adjoint formulation.

## 1 Introduction

The optimization of aerodynamic designs, from aircraft wings to vehicle bodies, relies heavily on the ability to evaluate the effect of design changes on performance metrics such as lift, drag, and overall efficiency. In recent years, adjoint methods [1] [2] [3] have established themselves as a powerful tool for solving these complex optimization problems, enabling efficient gradient-based optimization with respect to a wide range of design variables. These methods, which leverage the adjoint equations derived from the governing physical equations, allow for the computation of sensitivities with minimal computational overhead.

While adjoint methods have been widely applied in the context of compressible, viscous flow governed by the Navier-Stokes equations, their application to potential flow models –which describe inviscid, irrotational flow–, starting with Jameson's seminal work [3], offers unique advantages in certain aerodynamic problems. Potential flow can offer comparable accuracy to Euler solvers at a significantly reduced cost, making them an attractive starting point for aerodynamic optimization, especially in early-stage design or scenarios where computational resources are limited.

In this paper, we will focus on the adjoint formulation for incompressible potential flows. Most published work has focused on compressible potential flows, see [3] [4] [5] [6] [7] [8] [9] [10] [11], and references therein. Application to incompressible potential flow can be found, for example, in [11] [12] [13] [14]. While the former deals with the adjoint of the full potential equation, the latter focuses on the adjoint of Laplace's equation. This is related, but different, to the case addressed in [15], which studied the adjoint incompressible Euler equations in 2D when the base flow is irrotational. One of the goals of this work is precisely to establish the relation between both approaches.

The adjoint of Laplace's equation is itself a Laplace equation with modified boundary conditions related to the cost function. For simple cases (i.e., flow around a circle), the adjoint problem is exactly solvable in principle, but the results, when the cost function is the aerodynamic lift, are wrong. This result has been known to the authors for some time now, but its significance was not fully appreciated until recently. In deriving the analytic lift-based adjoint solution to the incompressible Euler equations in [15], a similar problem was encountered, which could only be resolved by explicitly taking into account the perturbation to the Kutta condition, resulting in a series of singularities in the adjoint solution even for blunt geometries. In what follows, we will see that the results of [15] can be used to obtain an analytic adjoint solution for incompressible potential flow, which can then be compared with the direct solution of the Laplace's equation to gain understanding of the cause of the problem.

One final issue that can be addressed with the analytic solutions is that of the Kutta and wake conditions. These are extremely relevant in potential flows, as they are required to select the physically meaningful solution, and their importance extends to adjoint potential solutions, which are meaningless without properly accounting for perturbations to the Kutta/wake conditions. While several approaches have been used to produce numerical adjoint solutions (see [10] [11] [12] for some recent examples with further references), continuous adjoint formulations for potential flow dealing with Kutta/wake boundary conditions are scarce (see however [4] [6] [9]), and a well-defined formulation is still lacking [11].

The relevance of these analysis is threefold. First, they provide with non-trivial analytic solutions which can be used not only to achieve a deeper understanding of the problem at hand, but can also serve as benchmark cases for adjoint potential solvers. A handful of analytic adjoint solutions are known. For quasi-1D inviscid flows, the Green's function approach has provided a tool to generate exact adjoint solutions [16] [17] in terms of the base flow solution. In inviscid 2D/3D flows, the entropy variables provide another exact solution for an output measuring net entropy flux across boundaries [18] [19]. A closely related solution, corresponding to a nearfield computation of aerodynamic drag, has been recently discovered by the authors [20], who have also provided a closed-form solution for incompressible inviscid flows in 2D [15] and outlined the general structure of the analytic solutions for subcritical compressible flows in [21]. Finally, exact solutions for the adjoint Navier-Stokes equations have been proposed for both laminar [22] and turbulent [23] boundary layer flows.

Second, they will certainly serve as a step towards understanding the compressible case whose basic form, for the Euler equations in subcritical flow, was outlined in [21]. Finally, they will also help to advance towards a potential continuous adjoint formulation including wake and Kutta conditions. Unraveling this issue will enable adjoint consistent discretizations, which bring about improved convergence properties [24] and are instrumental for error estimation.

## 2 Formulation of the adjoint potential equations

Consider the case of steady two-dimensional incompressible inviscid flow on a domain $\Omega$ with far-field boundary $S^{\infty}$ and wall boundary $S$ (typically an airfoil profile). In this set-up we are interested in the adjoint problem for a cost function measuring the non-dimensional aerodynamic force on a body of surface $S$ along a direction $\vec{d}$,

$$I = \int_S c_\infty^{-1}(p - p_\infty)(\hat{n}\cdot\vec{d})ds = \int_S c_\infty^{-1} p(\hat{n}\cdot\vec{d})ds \tag{1}$$

where $p$ is the pressure, $\hat{n}$ is the unit normal to $S$ pointing towards the interior of $S$ and $c_\infty = \rho q_\infty^{2}\ell/2$ is a normalization constant, with $\rho$ the fluid density, $q_\infty$ the free-stream velocity and $\ell$ a reference length (equal to the chord length for an airfoil or to twice the radius for a circle).

In order to set up the adjoint problem, we need to add to (1) the flow equations weighted by a Lagrange multiplier (the adjoint variable). Let us suppose that the flow is also irrotational. A two-dimensional steady irrotational flow can be described either by its potential $\phi$ or by the stream function, $\psi$, which in the case of an incompressible fluid are connected by the Cauchy-Riemann equations so that $\phi + i\psi$ is an analytic function, the complex potential, of a complex variable. Taking the real and imaginary part of the complex potential we obtain $\phi$ and $\psi$, respectively. Accordingly, there are a priori two ways to define the adjoint problem by considering perturbations to the potential or the stream function.

### 2.1 Adjoint Potential

In this first scenario, the velocity vector $\vec{v}$ is the gradient of a potential $\phi$, $\vec{v} = \nabla\phi$. The potential flow equation is

$$\nabla \cdot \vec{v} = \nabla^2\phi = 0 \tag{2}$$

with boundary conditions $\partial_n \phi_S = 0$ at the wall, where $\partial_n$ is the normal derivative, and $\phi \to q_\infty \vec{x} \cdot (\cos\alpha, \sin\alpha)$ at infinity, where $\alpha$ (the angle of attack) sets the flow direction far from the body. The Lagrangian can thus be written as

$$L = \int_S c_\infty{}^{-1} p(\hat{n} \cdot \vec{d})ds - \int_\Omega \tilde{\phi}(\nabla \cdot \vec{v})d\Omega = \int_S c_\infty{}^{-1} p(\hat{n} \cdot \vec{d})ds - \int_\Omega \tilde{\phi}\nabla^2\phi d\Omega \tag{3}$$

where $\tilde{\phi}$ is the adjoint potential and the pressure and the velocity are linked by Bernouilli's equation

$$p + \rho\vec{v} \cdot \vec{v} / 2 = \text{constant} \tag{4}$$

In order to derive the adjoint equations and boundary conditions, it suffices to linearize (3) with respect to flow variations $\delta\phi$ and $\delta p$. We will thus ignore the effect of shape variations, as these only contribute to the sensitivities and not to the adjoint problem formulation. We will also ignore the contribution of the Kutta condition and the wake cut (the discontinuity in the potential that arises in lifting bodies and which manifests itself as a non-zero monodromy of the potential when encircling the lifting body, $\Delta\phi = \pm\Gamma$, where $\Gamma$ is the circulation), see for example [6] [9]. The resulting variation in the Lagrangian is thus

$$\delta L = \int_S c_\infty{}^{-1} \delta p(\hat{n} \cdot \vec{d})ds - \int_\Omega \tilde{\phi}\nabla^2\delta\phi d\Omega \tag{5}$$

According to eq. (4), the pressure variation can be obtained in terms of the velocity variation as $\delta p = -\rho\vec{v} \cdot \delta\vec{v} = -\rho\nabla\phi \cdot \nabla\delta\phi$. Likewise, $\partial_n \phi_S = 0$ on *S*, so we can write $\delta p_S = -\rho\partial_s\phi\partial_s\delta\phi$ where $\partial_s$ represents the tangent derivative along the boundary. We can thus write eq. (5) as

$$\delta L = -\int_S c_\infty{}^{-1} \rho\partial_s\phi\partial_s\delta\phi(\hat{n} \cdot \vec{d})ds - \int_\Omega \nabla^2\tilde{\phi}\delta\phi d\Omega + \int_{\partial\Omega} \partial_n\tilde{\phi}\delta\phi ds - \int_{\partial\Omega} \tilde{\phi}\partial_n\delta\phi ds \tag{6}$$

were we have integrated by parts in the domain integral. We now integrate by parts in the linearized cost function [3, 5] and separate the boundary integrals into wall and far-field components [9] such that (6) yields

$$\delta L = \int_S c_\infty^{-1} \rho \partial_s (\partial_s \phi (\hat{n}\cdot\vec{d}))\delta\phi ds - \int_\Omega \nabla^2 \tilde{\phi}\delta\phi d\Omega + \int_S \partial_n \tilde{\phi}\delta\phi ds - \int_S \tilde{\phi}\partial_n \delta\phi ds + \int_{S_\infty} \partial_n \tilde{\phi}\delta\phi ds - \int_{S_\infty} \tilde{\phi}\partial_n \delta\phi ds \tag{7}$$

We can read off from (7) the adjoint equation $\nabla^2\tilde{\phi} = 0$ and wall boundary condition

$$\partial_n \tilde{\phi}_S = -c_\infty^{-1}\rho\partial_s(\partial_s\phi(\hat{n}\cdot\vec{d})) \tag{8}$$

This is eq. (25) in [5]. In the derivation of eq. (7), we have overlooked the term

$$-c_\infty^{-1}\rho[\partial_s\phi(\hat{n}\cdot\vec{d})\delta\phi]_{s=0} = -c_\infty^{-1}\rho\partial_s\phi(\hat{n}\cdot\vec{d})[\delta\phi]_{s=0} \tag{9}$$

that results from the integration by parts along the wall boundary. This is the jump in the bracketed quantity when the boundary $S$ is circled once. Note that the factor $\partial_s\phi(\hat{n}\cdot\vec{d})$ is always continuous, even if $s = 0$ corresponds to a sharp trailing edge, because if $s = 0$ is a cusp, then both $\partial_s\phi$ and $\hat{n}$ are continuous, and if $s = 0$ has a non-zero wedge angle, then $\hat{n}$ is discontinuous but $\partial_s\phi = 0$ (Kutta condition). Additionally, eq. (9) is not well defined as its value depends on the starting point of the integration (that we have taken to be at $s = 0$), being zero if $s = 0$ is a stagnation point. At any rate, keeping this term gives rise to a boundary forcing term involving a pair of Dirac delta functions centered at the starting point and with opposite signs (see [9], where an analogous term is derived as a result of integration along the wake cut, and also section 5.2 below)

$$\partial_n\tilde{\phi}_S = \pm c_\infty^{-1}\rho(\partial_s\phi(\hat{n}\cdot\vec{d}))_{s=0}\delta(s) \tag{10}$$

where the upper sign (resp. lower) corresponds to the upper limit (resp. lower limit) of the integration. Each of these terms, taken alone, would give rise to solutions with opposite signs the combined effect of which, given the linearity of the equations, should cancel out.

### 2.2 Adjoint Stream function

The velocity vector can be also described by a stream function as $\vec{v} = (\partial_y\psi, -\partial_x\psi)$. The stream function obeys the equation $\nabla\times\vec{v} = \partial_x v - \partial_y u = -\nabla^2\psi = 0$ throughout the domain, and it attains a constant value at the wall $S$, so $\partial_s\psi_S = 0$. In this setup, the Lagrangian takes the form

$$L = \int_S c_\infty^{-1} p(\hat{n}\cdot\vec{d})ds + \int_\Omega \tilde{\psi}(\nabla\times\vec{v})d\Omega = \int_S c_\infty^{-1} p(\hat{n}\cdot\vec{d})ds - \int_\Omega \tilde{\psi}\nabla^2\psi d\Omega \tag{11}$$

where $\tilde{\psi}$ is the adjoint stream function. Linearizing (11) with respect to flow variations and integrating by parts yields

$$\begin{aligned}\delta L &= \int_S c_\infty^{-1}\delta p(\hat{n}\cdot\vec{d})ds - \int_\Omega \tilde{\psi}\nabla^2\delta\psi d\Omega = -\int_S c_\infty^{-1}\rho\nabla^\perp\psi\cdot\nabla^\perp\delta\psi(\hat{n}\cdot\vec{d})ds - \int_\Omega \tilde{\psi}\nabla^2\delta\psi d\Omega = \\ &-\int_S c_\infty^{-1}\rho\nabla\psi\cdot\nabla\delta\psi(\hat{n}\cdot\vec{d})ds - \int_\Omega \tilde{\psi}\nabla^2\delta\psi d\Omega = -\int_S c_\infty^{-1}\rho\partial_n\psi\partial_n\delta\psi(\hat{n}\cdot\vec{d})ds - \int_\Omega \nabla^2\tilde{\psi}\delta\psi d\Omega \\ &-\int_S \tilde{\psi}\partial_n\delta\psi d\Omega + \int_S \partial_n\tilde{\psi}\delta\psi d\Omega - \int_{S_\infty} \tilde{\psi}\partial_n\delta\psi d\Omega + \int_{S_\infty} \partial_n\tilde{\psi}\delta\psi d\Omega\end{aligned} \tag{12}$$

where $\nabla^{\perp} = (\partial_y, -\partial_x)$. We can read off (12) the adjoint equation $\nabla^2\tilde{\psi} = 0$ and the wall boundary condition

$$\tilde{\psi}_S = -c_\infty{}^{-1}\rho\partial_n\psi(\hat{n}\cdot\vec{d}) = -c_\infty{}^{-1}\rho\nabla\psi\cdot\vec{d} = c_\infty{}^{-1}\rho(v,-u)\cdot\vec{d} \tag{13}$$

as well as $\tilde{\psi} \to 0$ on $S_\infty$.

It is relatively simple to write down a tentative solution for $\tilde{\psi}$. Since eq. (13) is written in terms of harmonic functions (the velocity components), we can extend it to the whole domain,

$$\tilde{\psi} = c_\infty{}^{-1}\rho(v,-u)\cdot\vec{d} + f \tag{14}$$

where $f$ is an arbitrary harmonic function vanishing on $S$. Eq. (14) obeys Laplace's equation and the wall boundary condition for any $\vec{d}$, but problems arise in the farfield. If $\vec{d} = (\cos\alpha, \sin\alpha)$, which corresponds to aerodynamic drag, Eq. (14), with $f$ = 0 everywhere, obeys the far-field boundary condition

$$\begin{aligned}&\tilde{\psi}_D = c_\infty{}^{-1}\rho(v,-u)\cdot(\cos\alpha,\sin\alpha) \to 0 \quad \text{on } S_\infty\\ &\text{since } (v,-u) \to q_\infty(\sin\alpha,-\cos\alpha) \quad \text{on } S_\infty\end{aligned}$$

If, on the other hand, $\vec{d} = (-\sin\alpha, \cos\alpha)$, which corresponds to lift,

$$\begin{aligned}&\tilde{\psi}_L = f_L + c_\infty{}^{-1}\rho(v,-u)\cdot(-\sin\alpha,\cos\alpha) \to f_L^\infty + c_\infty{}^{-1}\rho q_\infty(\sin\alpha,-\cos\alpha)\cdot(-\sin\alpha,\cos\alpha) =\\ &f_L^\infty - c_\infty{}^{-1}\rho q_\infty \quad \text{on } S_\infty\end{aligned}$$

The farfield adjoint b.c. thus requires $f_L^\infty = c_\infty{}^{-1}\rho q_\infty$. So $f$ must vanish on $S$ and attain a constant, non-zero value $f_L^\infty$ at infinity. This is impossible for "normal" functions. To see it, suppose that $S$ is a circle and consider a conformal transformation that maps the exterior of the circle to the interior, such that the farfield gets mapped into the origin. By a well-known feature of conformal transformations, $f$ expressed in the new coordinates is also a solution to Laplace's equation vanishing at the circle and attaining a non-zero value $f_L^\infty$ at the origin. But since harmonic functions obey the mean value property, which says that the value the function at the origin is given by its average value on the circle, the value of $f$ at the origin must be 0, which is a contradiction. We thus find an inconsistent solution for the naïve lift-based adjoint potential equations that we will unravel in section 3.

### 2.3 Relation of adjoint potential and stream function

As explained above, the potential and stream functions are the real and imaginary parts of a holomorphic function (the complex potential). As such, they are harmonic (obey the Laplace equation) and are linked by the Cauchy-Riemann equations

$$\frac{\partial\phi}{\partial x} = \frac{\partial\psi}{\partial y}, \ \frac{\partial\phi}{\partial y} = -\frac{\partial\psi}{\partial x} \tag{15}$$

It is reasonable to suspect that the adjoint potential and stream function are related in a similar fashion. For $\phi$ and $\psi$ the origin of eq. (15) is clear, since their derivatives yield the velocity components by construction, but the same is not immediately obvious for their duals. However, it turns out that $\tilde{\psi}+i\tilde{\phi}$ is a holomorphic function (so the roles are reversed with respect to the flow potential and stream functions). In order to prove it, we need to formulate the adjoint problem in terms of the velocities and not the potential/stream function. If we do that, we need to impose both the incompressibility and irrotationality conditions, i.e., the Lagrangian must contain both adjoint-weighted terms

$$L=\int_S c_\infty^{-1} p(\hat{n}\cdot\vec{d})ds-\int_\Omega \tilde{\phi}(\nabla\cdot\vec{v})d\Omega+\int_\Omega \tilde{\psi}(\nabla\times\vec{v})d\Omega \tag{16}$$

The sign in front of the adjoint stream function term in (16) has been introduced to conform to the sign conventions used in (3) and (11). Linearizing (16) with respect to velocity variations and integrating by parts we get

$$\begin{aligned}\delta L=&-c_\infty^{-1}\rho\int_S \vec{v}\cdot\delta\vec{v}(\hat{n}\cdot\vec{d})ds+\int_\Omega \nabla\tilde{\phi}\cdot(\delta u,\delta v)d\Omega-\int_\Omega(-\partial_y\tilde{\psi},\partial_x\tilde{\psi})\cdot(\delta u,\delta v)d\Omega\\&-\int_S\tilde{\phi}\hat{n}\cdot(\delta u,\delta v)ds+\int_S\tilde{\psi}\hat{n}\cdot(\delta v,-\delta u)ds-\int_{S_\infty}\tilde{\phi}\hat{n}\cdot(\delta u,\delta v)ds+\int_{S_\infty}\tilde{\psi}\hat{n}\cdot(\delta v,-\delta u)ds\end{aligned} \tag{17}$$

From (17) we get the coupled adjoint equations $\nabla\tilde{\phi}=(-\partial_y\tilde{\psi},\partial_x\tilde{\psi})$, which are the Cauchy-Riemann equations for the complex function $\tilde{\psi}+i\tilde{\phi}$.

### 2.4 Kutta condition

Potential flows carry an arbitrary amount of circulation that needs to be specified additionally and independently of the free-stream conditions (incidence and speed). For geometries with sharp corners, circulation is fixed by demanding that the flow be free of singularities at the corners . For regular geometries, an additional constraint (such as zero lift or a prescribed rotation velocity) is required. At any rate, such condition needs to be incorporated into the adjoint procedure, even for blunt bodies as demonstrated in [15]. We will see below that ignoring the effect of the Kutta condition leads to a wrong solution for the lift-based adjoint potential solution. It is then mandatory to incorporate the Kutta condition in the adjoint potential case. This is particularly critical when using the flow potential, which has a jump equal in value to the flow circulation when the airfoil is circled once. This is conventionally addressed by defining a cut line that is traditionally taken to extend from the trailing edge (t.e.) or rear stagnation point (r.s.p.) to the far-field. How this cut line and the associated trailing edge condition are treated in adjoint implementations differ among authors and, especially, between continuous [3] [4] [5] [9] and discrete [10] [12] adjoint methodologies. Jameson [3] [5] does not discuss the issue, while Reuther [6] addresses it in some detail but comes to the conclusion that no modifications are required as long as the circulation remains fixed. When the circulation is allowed to change, it was shown in [9] that a Dirac delta function is necessary as a boundary forcing term in the adjoint potential boundary condition. In [9], the Kutta condition is incorporated by equalizing the jump of the potential and the circulation both at the trailing edge and the wake cut, and also as a condition at infinity. The result is that the Dirac delta term in the adjoint boundary condition involves an integration of the adjoint variable along the cut and the far-field. This results in an implicit b.c. that, unfortunately, is actually not obeyed by the analytic adjoint solution that we will obtain below. Furthermore, no wake cut appears when using the stream function as the flow variable, so it is unclear how to deal with that issue in that particular case. In terms of the stream function, the Kutta condition can be imposed as [25]

$$(\hat{n}\cdot\nabla\psi)_u+(\hat{n}\cdot\nabla\psi)_l=0 \tag{18}$$

at the t.e. or r.s.p., expressing the assumption that the averaged (tangent) velocity between the upper and lower sides of the t.e./r.s.p. is zero; thus, the velocity in the vicinity of the t.e./r.s.p. is finite or zero depending on the details of the geometry. (An analog condition on the tangent derivatives of the potential at the trailing edge is used in [12] to impose the Kutta condition). A more convenient condition is obtained by using a conformal transformation to map the geometry to a circle. In the circle plane, the Kutta condition is

$$\hat{n}\cdot\nabla\psi\big|_{r.s.p.}^{circle}=0 \tag{19}$$

which simply means that the velocity vanishes at the r.s.p. of the circle. This needs to be true of any perturbation to the stream function, so the linearized condition would be

$$(\delta\vec{x}\cdot\nabla)(\nabla\psi)\cdot\hat{n}+\delta\hat{n}\cdot\nabla\psi+\hat{n}\cdot\nabla\delta\psi\big|_{r.s.p.}=0 \tag{20}$$

where $\delta\vec{x}$ is the displacement of the r.s.p. If the geometry remains fixed, eq. (20) reduces to

$$\hat{n}\cdot\nabla\delta\psi\big|_{r.s.p.}=0 \tag{21}$$

Under a conformal transformation,

$$\hat{n}\cdot\nabla\psi\big|_{r.s.p.}^{circle}=h\,\hat{n}\cdot\nabla\psi\big|_{te}^{airfoil} \tag{22}$$

where $h$ is the modulus of the conformal map $h=\left|\partial z/\partial\zeta\right|$ and $z$ and $\zeta$ are the complex coordinates in the airfoil and circle planes, respectively. This results in an (arguably) more convenient formulation of the perturbed Kutta condition

$$h\,\hat{n}\cdot\nabla\delta\psi\big|_{r.s.p.}^{airfoil}=h\,\partial_n\delta\psi\big|_{te}^{airfoil}=0 \tag{23}$$

The reason for the presence of $h$ can be further explained by taking into account that $\partial_n\psi$ or $\partial_n\delta\psi$ blow up near the trailing edge as $(\text{regular terms})+\kappa/r^{1-\pi/(2\pi-\tau)}$ [26], where $r$ is the distance to the trailing edge, $\tau$ is the trailing edge angle and $\kappa$ is a constant that depends on the circulation and that vanishes when the Kutta condition is enforced. Within the framework of conformal mapping, this behavior comes from the conformal transformation that turns the airfoil into a circle and whose modulus behaves as $h\sim r^{1-\pi/(2\pi-\tau)}$ (recall that the velocities in the airfoil plane behave as $1/h$). It is thus clear that $h$ in eq. (22) or (23) cancels the singularity of the velocity such that $h\,\partial_n\delta\psi\big|_{te}^{airfoil}\sim\kappa$.

We can add eq. (23) to the Lagrangian (11) with a new Lagrange multiplier $\tilde{\psi}_{te}$ as

$$\tilde{\psi}_{te}h\,\partial_n\psi\big|_{te}^{airfoil}=\int_S\tilde{\psi}_{te}h\partial_n\psi\delta(s-s_{te})ds \tag{24}$$

Eq. (24) introduces in the adjoint boundary condition (13) the additional contribution

$$\tilde{\psi}_{te}h\delta(s-s_{te}) \tag{25}$$

Unfortunately, neither the perturbed cost function nor the integral of the field equations produce, *a priori,* any term that can be matched with the above, so the value of the adjoint Kutta variable $\tilde{\psi}_{te}$ remains undefined. However, we will see in section 6 that there is a particular value of $\tilde{\psi}_{te}$ such that eq. (25) produces the correct boundary behavior compatible with the analytic solution.

## 3 Analytic solution of the adjoint Laplace's equation via Green's functions

Green's functions can be used to build analytic solutions for adjoint equations [15] [27] [16]. The adjoint state can be obtained as the linearized cost function due to a point perturbation. For potential flow, we have two formulations and, thus, there are two Green's functions that can be considered.

### 3.1 Green's function for the adjoint stream function

We can first consider the stream function formulation, eq. (11). Adding a point perturbation to the linearized flow equation yields

$$\delta L = \int_S c_\infty^{-1} \delta p (\hat{n}\cdot\vec{d}) ds + \int_\Omega \tilde{\psi}(\nabla\times\delta\vec{v} + \varepsilon\delta(\vec{x}-\vec{x}_0)) d\Omega \tag{26}$$

where $\delta(\vec{x}-\vec{x}_0)$ is the Dirac delta function and $\varepsilon$ is the strength of the perturbation that we assume to be small. With homogeneous boundary conditions for $\delta\vec{v}$ and provided that $\tilde{\psi}$ obeys the adjoint equation and boundary conditions (13), eq. (26) yields

$$\delta L = \varepsilon\tilde{\psi}(\vec{x}_0) \tag{27}$$

from where we can obtain the value of $\tilde{\psi}(\vec{x}_0)$ if we can compute the value of $\delta L$, which is the additional force exerted by the fluid on the body due to the presence of the point perturbation. This is the essence of the Green's function approach introduced by Giles and Pierce [16].

Now notice that the perturbation obeys the equation

$$\nabla\times\delta\vec{v} = (\partial_x\delta v - \partial_y\delta u) = -\varepsilon\delta(\vec{x}-\vec{x}_0) \tag{28}$$

This is a point vortex, which in free space is described by the complex potential [28]

$$\Phi(z) = \frac{i\varepsilon}{2\pi}\ln(z-z_0) \tag{29}$$

where $z = x+iy$ is a complex coordinate and $z_0$ represents the complex coordinate of the center of the vortex.

The linearized force exerted by the vortex on $S$ was computed in [15]. For a vortex located at $\vec{x}_0$, the complex force $\delta D - i\delta L$ (where $\delta D$ and $\delta L$ are the linearized drag and lift forces on $S$) is

$$\delta D - i\delta L = i\rho q_\infty \delta\Gamma_0 + i\rho\varepsilon e^{i\alpha}(u - iv - q_\infty e^{-i\alpha})_{\vec{x}=\vec{x}_0} + O(\varepsilon^2) \tag{30}$$

Here, $\delta\Gamma_0$ is a perturbed circulation that comes from the perturbation to the Kutta condition: the vortex perturbs the flow at the trailing edge or rear stagnation point; $\delta\Gamma_0$ is the amount of circulation required to restore the Kutta condition. By using a conformal mapping $\zeta(z)$ of the profile to a circle, it was found in [15] that

$$\delta\Gamma_0 = -\varepsilon\Upsilon^{(2)}(\zeta(z_0)) \tag{31}$$

(we are using conventions such that $\delta\Gamma_0 > 0$ corresponds to a counter-clockwise vortex) where

$$\Upsilon^{(2)}(\zeta(z_0)) = \left( \frac{\zeta_{te} - \zeta_c}{\zeta_0 - \zeta_{te}} + \frac{\bar{\zeta}_{te} - \bar{\zeta}_c}{\bar{\zeta}_0 - \bar{\zeta}_{te}} \right) \tag{32}$$

In eq. (32), $\zeta$ and $z$ are the complex coordinates in the planes of the circle and the profile, respectively, $\zeta_c$ is the position of the circle center and $\zeta_{te}$ is the location of the rear stagnation point on the circle.

From eqs. (30)-(32), the linearized lift and drag coefficients read

$$\begin{aligned} \delta(c_D - ic_L)_{\vec{x}_0} &= \frac{i\rho}{c_\infty}(q_\infty \delta\Gamma_0 + \varepsilon e^{i\alpha}(u - iv - q_\infty e^{-i\alpha})) = \\ &\frac{\rho\varepsilon}{c_\infty}(v\cos\alpha - u\sin\alpha + i(u\cos\alpha + v\sin\alpha - q_\infty(1+\Upsilon^{(2)}))) \end{aligned} \tag{33}$$

Since $\delta(c_D - ic_L)_{\vec{x}_0} = \varepsilon(\tilde{\psi}_D(\vec{x}_0) - i\tilde{\psi}_L(\vec{x}_0))$, the adjoint stream functions for drag and lift are given by the drag and lift exerted by the point vortex on the body

$$\begin{aligned} \tilde{\psi}_D(\vec{x}_0)) &= \frac{\rho}{c_\infty}(v\cos\alpha - u\sin\alpha)_{\vec{x}=\vec{x}_0} \\ \tilde{\psi}_L(\vec{x}_0) &= -\frac{\rho}{c_\infty}(u\cos\alpha + v\sin\alpha - q_\infty(1+\Upsilon^{(2)}))_{\vec{x}=\vec{x}_0} \end{aligned} \tag{34}$$

It is interesting to recall at this point the analysis performed in section 2.2. We found there the tentative solution $\tilde{\psi} = c_\infty^{-1}\rho(v,-u)\cdot\vec{d} + f$, which agrees with (34) if $f = 0$ for drag and $f = \rho q_\infty(1+\Upsilon^{(2)})/c_\infty$ for lift , where

$$1+\Upsilon^{(2)} = 1 + \frac{\zeta_{te} - \zeta_c}{\zeta_0 - \zeta_{te}} + \frac{\bar{\zeta}_{te} - \bar{\zeta}_c}{\bar{\zeta}_0 - \bar{\zeta}_{te}} = \frac{|\zeta_0 - \zeta_c|^2 - |\zeta_{te} - \zeta_c|^2}{|\zeta_0 - \zeta_{te}|^2} \tag{35}$$

is the Poisson kernel for the circle [29] (the Poisson kernel is an integral kernel used for solving the 2D Laplace equation on a disk with Dirichlet boundary conditions, and can be obtained as the derivative of the Green's function for the Laplacian). The Poisson kernel vanishes at the circle and attains the value 1 at infinity, but it also has a singularity at $\zeta_{te}$ and, in fact, it behaves as a Dirac delta function $\delta(\zeta_{te})$ on the circle [29]. This is the loophole that allows an *f* to exist despite what we argued in section 2.2, since $1+\Upsilon^{(2)}$ is not a "normal" function but a Dirac delta distribution: it is zero almost everywhere on the circle but its average is 1.

### 3.2 Green's function for the adjoint potential

We next consider the potential formulation given by eq. (3). Adding a point perturbation to the linearized flow equation we get

$$\delta L = \int_S c_\infty^{-1}\delta p(\hat{n}\cdot\vec{d})ds - \int_\Omega \tilde{\phi}(\nabla\cdot\delta\vec{v} - \varepsilon\delta(\vec{x}-\vec{x}_0))d\Omega \tag{36}$$

With homogeneous boundary conditions for $\delta\vec{v}$ and provided that $\tilde{\phi}$ obeys Laplace's equation $\nabla^2\tilde{\phi} = 0$ and the wall boundary condition $\partial_n\tilde{\phi}_S = -c_\infty^{-1}\rho\partial_t(\partial_t\phi(\hat{n}\cdot\vec{d}))$, eq. (36) yields

$$\delta L = \varepsilon \tilde{\phi}(\vec{x}_0) \tag{37}$$

from where we can obtain the value of $\tilde{\phi}(\vec{x}_0)$ in terms of $\delta L$.

The perturbation obeys the equation

$$\nabla \cdot \delta \vec{v} = \varepsilon \delta(\vec{x} - \vec{x}_0) \tag{38}$$

This is a point (mass) source, which in free space is described by the complex potential

$$\Phi(z) = \frac{\varepsilon}{2\pi} \ln(z - z_0) \tag{39}$$

The linearized lift and drag coefficients exerted by the point source are [15]

$$\begin{aligned}\delta(c_D - ic_L)_{\vec{x}_0} &= \frac{\rho}{c_\infty}(iq_\infty \delta\Gamma_0 + \varepsilon e^{i\alpha}(u - iv - q_\infty e^{-i\alpha})) = \\ &= \frac{\rho\varepsilon}{c_\infty}(u\cos\alpha + v\sin\alpha - q_\infty + i(u\sin\alpha - v\cos\alpha - q_\infty \Upsilon^{(1)}))\end{aligned} \tag{40}$$

where

$$\delta\Gamma_0 = -\varepsilon \Upsilon^{(1)}(\zeta(z_0)) \tag{41}$$

and

$$\Upsilon^{(1)}(\zeta(z_0)) = -i\left(\frac{\zeta_{te} - \zeta_c}{\zeta_0 - \zeta_{te}} - \frac{\bar{\zeta}_{te} - \bar{\zeta}_c}{\bar{\zeta}_0 - \bar{\zeta}_{te}}\right) \tag{42}$$

is the conjugate Poisson kernel. Since $\delta(c_D - ic_L)_{\vec{x}_0} = \varepsilon(\tilde{\phi}_D(\vec{x}_0) - i\tilde{\phi}_L(\vec{x}_0))$, the analytic adjoint potential solutions for drag and lift are given by the drag and lift exerted by the point source

$$\begin{aligned}\tilde{\phi}_D(\vec{x}_0) &= \frac{\rho}{c_\infty}(u\cos\alpha + v\sin\alpha - q_\infty)_{\vec{x}=\vec{x}_0} \\ \tilde{\phi}_L(\vec{x}_0) &= \frac{\rho}{c_\infty}(v\cos\alpha - u\sin\alpha + q_\infty \Upsilon^{(1)})_{\vec{x}=\vec{x}_0}\end{aligned} \tag{43}$$

Again, the drag-based adjoint potential is smooth throughout the flow field, but the lift-based solution has a singularity at the trailing edge (rear stagnation point in the case of the circle) due to $\Upsilon^{(1)}$.

### 3.3 Adjoint of circulation

The Kutta functions $\Upsilon^{(1)}$ and $\Upsilon^{(2)}$ turn out to have a simple interpretation beyond their being the circulation required to restore the Kutta condition or their mathematical identification as the Poisson kernel and its harmonic conjugate. They correspond to the analytic solution for the adjoint potential and stream function associated to the cost function $\int_S \vec{v} \cdot d\vec{l}$ that measures the circulation around the body.

The analytic solution for the corresponding adjoint problems can be obtained using the circulation $\delta\Gamma_0$ added by a point source or vortex, which is entirely due to the Kutta condition. Recalling the results in section 3, we have

$$\begin{aligned}\tilde{\phi} &= -\Upsilon^{(1)} \\ \tilde{\psi} &= -\Upsilon^{(2)}\end{aligned} \tag{44}$$

so the solution in this case contains only the singular functions.

## 4 Relation of potential adjoint and incompressible adjoint formulations

We can also obtain the analytic adjoint solutions for potential flow from the analytic solution to the incompressible Euler equations obtained in [15]. This approach has the added benefit of unveiling the relation between both adjoint problems, which can be used to shed light on some features of the solution discussed in [15] and to sharpen the conjecture for the compressible case presented in [21].

In order to be as self-contained as possible, we begin by recalling a few facts regarding the inviscid (incompressible) adjoint equations. The primal flow is governed by the incompressible Euler equations $R(U) = \nabla \cdot \vec{F}(U) = \partial_x F_x + \partial_y F_y = 0$, where

$$U = \begin{pmatrix} p \\ \rho u \\ \rho v \end{pmatrix},\; F_x = \begin{pmatrix} \rho u \\ \rho u^2 + p \\ \rho uv \end{pmatrix},\; F_y = \begin{pmatrix} \rho v \\ \rho vu \\ \rho v^2 + p \end{pmatrix} \tag{45}$$

To obtain the sensitivities of the cost function (1) with respect to flow perturbations $\delta U$ using the adjoint approach we start by forming the Lagrangian

$$L = \int_S c_\infty^{-1}(p - p_\infty)(\hat{n}\cdot\vec{d})ds - \int_\Omega \psi^T \nabla \cdot \begin{pmatrix} \rho\vec{v} \\ \rho u\vec{v} + p\hat{x} \\ \rho v\vec{v} + p\hat{y} \end{pmatrix} d\Omega \tag{46}$$

where $\psi = (\psi_1, \psi_2, \psi_3)^T$ are the adjoint variables. We now assume, as in [15], that the flow is irrotational. Hence, $p + \rho\vec{v}^2/2$ is constant, so that we can write Eq. (46) as

$$\begin{aligned} L &= \int_S c_\infty^{-1}(p - p_\infty)(\hat{n}\cdot\vec{d})ds - \int_\Omega \rho\psi^T \begin{pmatrix} 1 \\ u \\ v \end{pmatrix} \nabla\cdot\vec{v} d\Omega - \int_\Omega \rho\psi^T \begin{pmatrix} 0 \\ -v \\ u \end{pmatrix} \nabla\times\vec{v} d\Omega = \\ &\int_S c_\infty^{-1}(p - p_\infty)(\hat{n}\cdot\vec{d})ds - \int_\Omega \rho(\psi_1 + u\psi_2 + v\psi_3)(\nabla\cdot\vec{v})d\Omega + \int_\Omega \rho(v\psi_2 - u\psi_3)(\nabla\times\vec{v})d\Omega \end{aligned} \tag{47}$$

Hence, and by comparison of eq. (47) with eq. (16), we find the following correspondence between the potential and Euler incompressible adjoint states

$$\begin{aligned}\tilde{\phi} &= \rho(\psi_1 + u\psi_2 + v\psi_3) \\ \tilde{\psi} &= \rho(v\psi_2 - u\psi_3)\end{aligned} \tag{48}$$

There are three points worth-mentioning concerning the above identifications. First, following [27] it was established in [15] that, for the 2D incompressible Euler equations, the combinations $I^{(1)} = \psi_1 + u\psi_2 + v\psi_3$ and $I^{(2)} = v\psi_2 - u\psi_3$ yield the linearized cost function due to the point mass source and the point vortex, respectively, so eq. (48) confirms the pairing of the adjoint potential with the point source and of the adjoint stream function with the point vortex that we unveiled in section 3.

Second, it is possible to use eq. (48) to derive the analytic solutions for the adjoint potential and stream function from the analytic solutions obtained in [15] for $(\psi_1, \psi_2, \psi_3)$. Using the analytic solutions of [15] (eq. 4.12) we obtain exactly Eqs. (43) and (34).

Finally, the relation sheds light on a curious behavior observed in [15]. The analytic lift-based adjoint solutions for the incompressible Euler equations contain a singularity along the dividing streamline upstream of the rear stagnation point (for the circle) or trailing edge (for airfoils), such that the adjoint variables are actually divergent along the wall. However, there are two combinations of adjoint variables, which are related to the continuous adjoint expression for the sensitivity derivatives and to the adjoint wall boundary condition, respectively, which have a finite value along the wall. These are, precisely, $I^{(1)}$ and $I^{(2)}$, which, as we have just seen, correspond to the solutions to the adjoint Laplace's equation. The origin of the streamline singularity is in fact in a third, non-potential perturbation $I^{(3)}$ that changes the total pressure of the flow.

## 5 Boundary behavior of the analytic adjoint solutions

In order to advance towards a formulation of the adjoint problem for potential flows we need to understand the behavior of the analytic solutions at wall boundaries. We have seen that the naïve approach does not lead to the correct solution even in the arguably simplest case involving the stream function. In what follows, we will examine the behavior of the analytic solutions at the boundary.

### 5.1 Adjoint stream function

The analytic adjoint stream function is given by eq. (34). Suppose that $\alpha = 0$ and that the wall boundary $S$ is a circle of radius $R$ centered at the origin. Then the rear stagnation point is at $(x, y) = (R, 0)$ and the lift-based adjoint stream function at the boundary takes the form

$$\begin{aligned} \tilde{\psi}_L\Big|_s &= \frac{\rho}{c_\infty}(v,-u)_S \cdot (-\sin\alpha, \cos\alpha) + \frac{\rho q_\infty}{c_\infty}(1+\Upsilon^{(2)}) = \\ &-\frac{\rho}{c_\infty}(v,-u)_S \cdot (-\sin\alpha, \cos\alpha) + \frac{\rho q_\infty}{c_\infty}\frac{r^2 - R^2}{r^2 + R^2 - 2rR\cos\theta} \end{aligned} \tag{49}$$

using polar coordinates $(r, \theta)$ with origin at the circle center. The function

$$P_{r,R}(\theta) = \frac{1}{2\pi}\frac{r^2 - R^2}{r^2 + R^2 - 2rR\cos\theta} \tag{50}$$

is the Poisson kernel for the exterior of a circle of radius $R$ [29]. The Poisson kernel obeys

$$P_{r,R}(\theta) \to \delta(\theta) \text{ as } r \to R \tag{51}$$

where $\delta(\theta)$ is the Dirac delta distribution, so it turns out that the Kutta function $1+\Upsilon^{(2)}$ corresponds to the solution of Laplace's equation with a Dirac delta function at the rear stagnation point. The correct wall boundary behavior is thus

$$\tilde{\psi}_L\Big|_s = \frac{\rho}{c_\infty}(v,-u)_S\cdot(-\sin\alpha,\cos\alpha)+\frac{2\pi\rho q_\infty}{c_\infty}\delta(\theta) \quad (52)$$

In the general case (an airfoil), the boundary stream function is still given by the first line of (49). Hence, the Kutta function is the same but expressed in the appropriate coordinates. i.e., on the airfoil plane ($z$) the Kutta function is

$$\Upsilon^{(2)}(z)=\frac{\zeta(z_{te})}{\zeta(z)-\zeta(z_{te})}+\frac{\bar{\zeta}(z_{te})}{\bar{\zeta}(z)-\bar{\zeta}(z_{te})} \quad (53)$$

where $\zeta(z)$ is the corresponding complex coordinate on the circle plane. Hence, $1+\Upsilon^{(2)}(z)$ is not the Poisson kernel on the airfoil plane $P_z$, but is related to it by multiplication by the modulus of the conformal mapping $h=\left|\partial z/\partial\zeta\right|$,

$$1+\Upsilon^{(2)}(z)=hP_z \quad (54)$$

From a physical point of view this is clear, since $\Upsilon^{(2)}$ is obtained by applying the Kutta condition on the circle plane. From a mathematical point of view, it is also natural to expect so, since $1+\Upsilon^{(2)}$ is essentially a Dirac delta and Dirac deltas behave in precisely that way under changes of variables. The conclusion is that the behavior for the boundary stream function on the airfoil plane is

$$\tilde{\psi}_L\Big|_s = \frac{\rho}{c_\infty}(v,-u)_S\cdot(-\sin\alpha,\cos\alpha)+\frac{2\pi\rho q_\infty}{c_\infty}Rh\delta(s) \quad (55)$$

where $s$ is the arc-length parameter on the airfoil defined in such a way that $s = 0$ corresponds to the location of the trailing edge.

### 5.2 Adjoint potential

The analytic solution for the adjoint potential (43) contains a singular term $c_\infty^{-1}\rho q_\infty\Upsilon^{(1)}$. Specializing again to the case of a circle at zero incidence

$$c_\infty^{-1}\rho q_\infty\Upsilon^{(1)}=-c_\infty^{-1}\rho q_\infty\frac{2Rr\sin\theta}{r^2+R^2-2rR\cos\theta} \quad (56)$$

Eq. (56) gives rise to the following term in the adjoint boundary conditions

$$\begin{aligned}&\partial_n\tilde{\phi}_S=(\text{regular part})-c_\infty^{-1}\rho q_\infty\,\partial_r\Upsilon^{(1)}\Big|_{r\to R}=\\&(\text{regular part})+c_\infty^{-1}\rho q_\infty R^{-1}\,\partial_\theta(1+\Upsilon^{(2)})\Big|_{r\to R}=\\&(\text{regular part})+\frac{2\pi\rho q_\infty}{c_\infty R}\partial_\theta\delta(\theta)\end{aligned} \quad (57)$$

where we have used that $\Upsilon^{(1)}$ and $\Upsilon^{(2)}$ are harmonic conjugates and, thus, $\partial_r\Upsilon^{(1)}=-\frac{1}{r}\partial_\theta(1+\Upsilon^{(2)})$. The Kutta function $\Upsilon^{(1)}$ thus corresponds to the solution of Laplace's equation with Neumann boundary conditions containing the derivative of a Dirac delta function at the rear stagnation point.

As in the previous case, the Kutta function does not change in the airfoil plane and taking into account the transformation properties of the Dirac delta, the behaviour of the adjoint potential at the boundary in the airfoil plane is given by

$$\partial_n \tilde{\phi}_S = (\text{regular part}) + \frac{2\pi\rho q_\infty R}{c_\infty}\partial_s (h\delta(s)) \tag{58}$$

It's instructive at this point to compare the above result with [9], who use a potential formulation for inverse design, and obtain a delta function in the adjoint boundary condition that comes from the jump in potential across the wake for lifting bodies,

$$\partial_n \tilde{\phi}\Big|_S = (\text{regular part}) \pm K\delta(x - x_{t.e.}) \tag{59}$$

where

$$K = \int_{cut} \partial_n \tilde{\phi} ds + \frac{1}{2\pi}\int_{S_\infty} \tilde{\phi}\partial_n \theta ds \tag{60}$$

and the upper/lower signs correspond to the lower/upper surfaces of the airfoil, respectively. This does not agree with eq. (58) and, besides it does not seem compatible with the analytic solution (43) . In fact, we have argued in 2.1 that the term $\pm K\delta(x - x_{t.e.})$ in the boundary condition does not give rise to any non-trivial solution.

## 6 On the Kutta condition for the adjoint equations

We have seen that the analytic adjoint solutions contain singular terms at the wall boundary involving the Dirac delta function and its derivative that we don't know a priori how to justify within the adjoint approach outlined in section 2. We argue here that a possible way out is to assume that the linearized lift contains an extra term localized at the trailing edge (or rear stagnation point) that comes from the perturbation to the Kutta condition.

*Stream function*

The singular term in the boundary adjoint stream function (55) can be obtained if the linearized Lagrangian (12) contains the term

$$\int_S 2\pi\rho q_\infty c_\infty^{-1} Rh\delta(s)\partial_n \delta\psi ds = 2\pi\rho q_\infty c_\infty^{-1} Rh\partial_n \delta\psi\Big|_{t.e.} \tag{61}$$

This looks much like the term introduced by the Kutta condition in (25) with a precise value for the adjoint Kutta variable $\tilde{\psi}_{te} = 2\pi\rho q_\infty c_\infty^{-1} R$. Eq. (61) raises two issues. The first one is that, as outlined in section 2.4, the unit normal vector at the sharp trailing edge is generally discontinuous, so the normal derivative is a priori meaningless. However,

$$h\partial_n \delta\psi\Big|_{airfoil} = \partial_n \delta\psi\Big|_{circle} \tag{62}$$

which is continuous at the rear stagnation point.

The second issue concerns, of course, the interpretation of (61). It turns out that the additional term admits a simple interpretation as enforcing the Kutta condition at the level of the linearized functional. The idea is as follows. The linearized cost function

$$\delta I = -\int_S c_\infty^{-1} \rho \vec{v} \cdot \delta \vec{v} (\hat{n} \cdot \vec{d}) ds \tag{63}$$

yields, for lift, $-c_\infty^{-1} \rho q_\infty \delta\Gamma_0 + ...$ where $\delta\Gamma_0$ is the perturbed circulation. The remaining terms are specific to the problem under consideration. For the point vortex, for example,

$$-\int_S c_\infty^{-1} \rho \vec{v} \cdot \delta \vec{v} (\hat{n} \cdot \vec{d}) ds = -c_\infty^{-1} \rho q_\infty \delta\Gamma_0 - c_\infty^{-1} \rho \varepsilon (u \cos\alpha + v \sin\alpha - q_\infty) \tag{64}$$

The value of $\delta\Gamma_0$ is fixed by imposing that the perturbed velocity at the sharp trailing edge be finite, which requires the perturbed velocity at the rear stagnation point of the circle, $\partial_n \delta\psi\big|_{r.s.p.}^{circle}$, to vanish. We thus have, recalling (62),

$$h \partial_n \delta\psi\big|_{t.e.}^{airfoil} = \partial_n \delta\psi\big|_{r.s.p.}^{circle} = \frac{\delta\Gamma_0}{2\pi R} + \delta v_t^{r.s.p.} = 0 \Rightarrow \delta\Gamma_0 = -2\pi R \delta v_t^{r.s.p.} \tag{65}$$

$\delta v_t^{r.s.p.}$ is the tangent perturbation velocity at the rear stagnation point excluding the circulation part. For the vortex, $\delta v_t^{r.s.p.}$ is the velocity induced by the vortex at the rear stagnation point, $\delta v_t^{r.s.p.} = \varepsilon \Upsilon^{(2)} / (2\pi R)$ [15] and, thus, the Kutta condition results in $\delta\Gamma_0 = -\varepsilon \Upsilon^{(2)}$.

On the other hand, if we do not impose the Kutta condition on the perturbation, $\delta\Gamma_0$ and, hence, $\delta I$ are undefined. We can still obtain the right value if the linearized lift is given by

$$\delta c_L = -\int_S c_\infty^{-1} \rho \vec{v} \cdot \delta \vec{v} (\hat{n} \cdot \vec{d}) ds + 2\pi \rho q_\infty c_\infty^{-1} R h \partial_n \delta\psi\big|_{t.e.} \tag{66}$$

since

$$\begin{aligned} \delta c_L &= -\int_S c_\infty^{-1} \rho \vec{v} \cdot \delta \vec{v} (\hat{n} \cdot \vec{d}) ds + 2\pi R c_\infty^{-1} \rho q_\infty h \partial_n \delta\psi\big|_{t.e.} = \\ & -c_\infty^{-1} \rho q_\infty \delta\Gamma_0 + ... + 2\pi R c_\infty^{-1} \rho q_\infty \partial_n \delta\psi\big|_{r.s.p.}^{circle} = \\ & -c_\infty^{-1} \rho q_\infty \delta\Gamma_0 + ... + 2\pi R c_\infty^{-1} \rho q_\infty \left( \frac{\delta\Gamma_0}{2\pi R} + \delta v_t^{r.s.p.} \right) = 2\pi R c_\infty^{-1} \rho q_\infty \delta v_t^{r.s.p.} + ... \end{aligned} \tag{67}$$

One can easily check now that (66) gives rise to the correct adjoint boundary condition (55).

*Potential*

In a similar fashion, the singular term in the boundary adjoint potential (58) can be obtained by assuming that the perturbed lift gets an extra term related to the Kutta condition. For the potential, the Kutta condition requires the perturbed velocity at the r.s.p. on the circle plane to vanish,

$$h \partial_t \delta\phi\big|_{t.e.}^{airfoil} = \partial_t \delta\phi\big|_{r.s.p.}^{circle} = 0 \tag{68}$$

(where $\partial_t$ is the tangent derivative along the boundary), which results in a finite velocity at the sharp trailing edge. Eq. (68) then sets the correct value of the perturbed circulation (65). As before, we can obtain the correct value of the perturbed lift without imposing (68) if the linearized lift is given by

$$\delta c_L = -\int_S c_\infty^{-1} \rho \vec{v} \cdot \delta \vec{v} (\hat{n} \cdot \vec{d}) ds + 2\pi R c_\infty^{-1} \rho q_\infty h \partial_t \delta\phi \Big|_{t.e.}^{airfoil} \tag{69}$$

since

$$\begin{aligned} &\delta c_L = -\int_S c_\infty^{-1} \rho \vec{v} \cdot \delta \vec{v} (\hat{n} \cdot \vec{d}) ds + 2\pi R c_\infty^{-1} \rho q_\infty h \partial_t \delta\phi \Big|_{t.e.}^{airfoil} = \\ &-c_\infty^{-1} \rho q_\infty \delta\Gamma_0 + \ldots + 2\pi R c_\infty^{-1} \rho q_\infty \partial_t \delta\phi \Big|_{t.e.}^{circle} = \\ &-c_\infty^{-1} \rho q_\infty \delta\Gamma_0 + \ldots + 2\pi R c_\infty^{-1} \rho q_\infty \left( \frac{\delta\Gamma_0}{2\pi R} + \delta v_t^{r.s.p.} \right) = 2\pi R c_\infty^{-1} \rho q_\infty \delta v_t^{r.s.p.} + \ldots \end{aligned} \tag{70}$$

For the point source, the perturbation velocity is $\delta v_t^{r.s.p.} = \varepsilon \Upsilon^{(1)} / (2\pi R)$ so Eq. (70) yields the correct value for the lift $\delta c_L = \varepsilon c_\infty^{-1} \rho q_\infty \Upsilon^{(1)} + \ldots$

Besides, recalling Eq. (7) and using eq. (69), the adjoint b.c. (58) is obtained from

$$\begin{aligned} &\int_S c_\infty^{-1} \rho \partial_t (\partial_t \phi (\hat{n} \cdot \vec{d})) \delta\phi ds + 2\pi R c_\infty^{-1} \rho q_\infty h \partial_t \delta\phi \Big|_{t.e.}^{airfoil} + \int_S \partial_n \tilde{\phi} \delta\phi ds = \\ &\int_S c_\infty^{-1} \rho \partial_t (\partial_t \phi (\hat{n} \cdot \vec{d})) \delta\phi ds + \int_S 2\pi R c_\infty^{-1} \rho q_\infty h \partial_s \delta\phi \delta(s) ds + \int_S \partial_n \tilde{\phi} \delta\phi ds = \\ &\int_S c_\infty^{-1} \rho \partial_t (\partial_t \phi (\hat{n} \cdot \vec{d})) \delta\phi ds - \int_S 2\pi R c_\infty^{-1} \rho q_\infty \partial_s (h \delta(s)) \delta\phi ds + \int_S \partial_n \tilde{\phi} \delta\phi ds = 0 \Rightarrow \\ &\partial_n \tilde{\phi} \Big|_S = -c_\infty^{-1} \rho \partial_t (\partial_t \phi (\hat{n} \cdot \vec{d})) + 2\pi R c_\infty^{-1} \rho q_\infty \partial_s (h \delta(s)) \end{aligned} \tag{71}$$

## 7 Sample solutions

In order to illustrate the results presented above, we present now the results of direct evaluation of the analytic adjoint solutions in two cases: a blunt body (a circle) and an airfoil with a sharp non-cusped trailing edge.

### 7.1 Circle

Figure 1 shows the adjoint potential and stream function for lift for incompressible potential flow around a circle, with zero incidence ($\alpha = 0°$) and circulation. The flow solution can be obtained with standard complex variable techniques [28] as

$$u = q_\infty \left( 1 - \frac{R^2}{r^2} \cos 2\theta \right), \; v = -q_\infty \frac{R^2}{r^2} \sin 2\theta \tag{72}$$

in terms of polar coordinates $(r, \theta)$ where $r$ is the distance from the center of the circle and $\theta$ is the angle with the positive $x$-axis. Using (34), (43) and (72), the lift-based analytic adjoint solutions are

$$\begin{aligned} \tilde{\phi}_L &= -\frac{\rho q_\infty}{c_\infty} \left( \frac{R^2}{r^2} \sin 2\theta + \frac{2Rr \sin\theta}{r^2 + R^2 - 2Rr\cos\theta} \right) \\ \tilde{\psi}_L &= \frac{\rho q_\infty}{c_\infty} \left( \frac{R^2}{r^2} \cos 2\theta + \frac{2R(r\cos\theta - R)}{r^2 + R^2 - 2Rr\cos\theta} \right) \end{aligned} \tag{73}$$

As seen in Figure 1, the adjoint solutions are singular at the rear stagnation point.

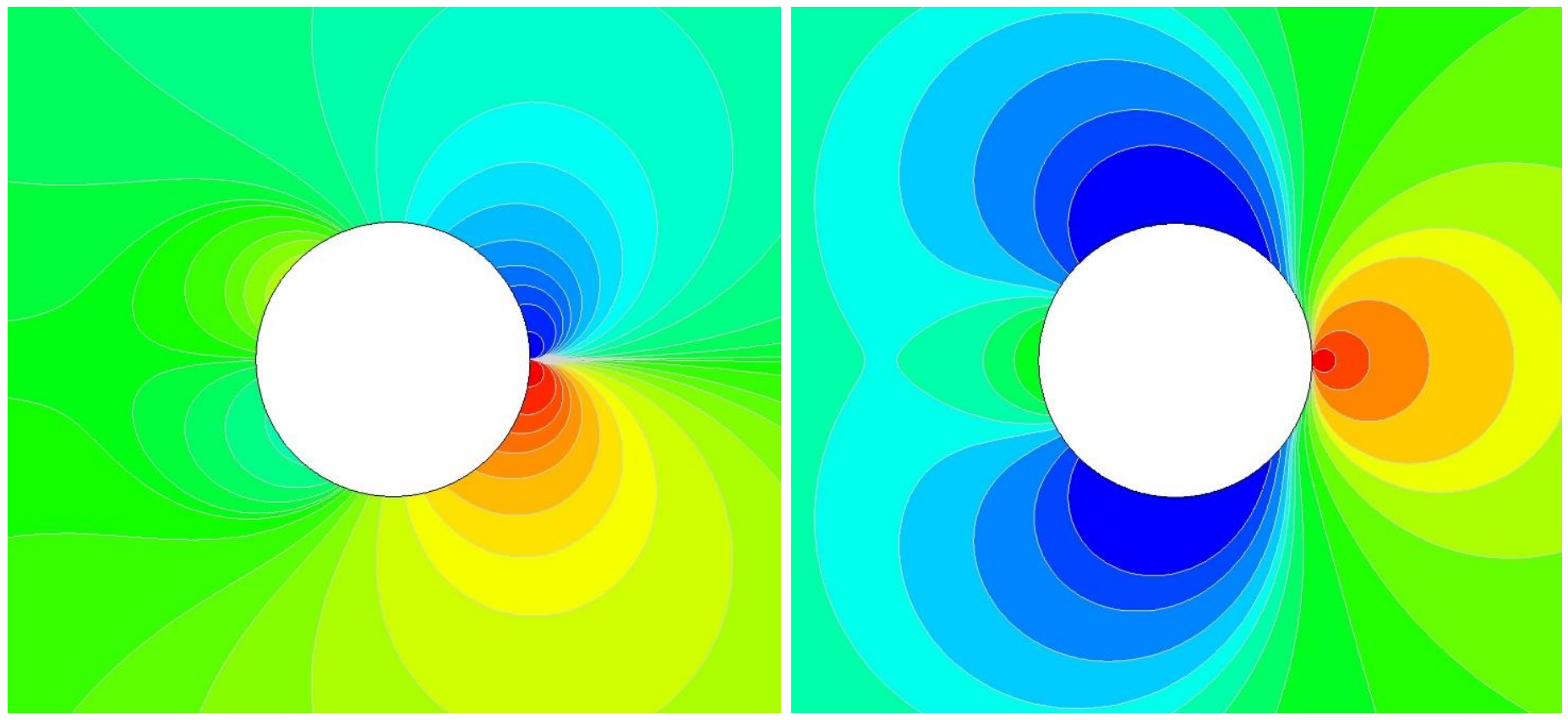

**Figure 1.** Analytic lift-based adjoint solution for incompressible potential flow around a circle. Left: adjoint potential $\tilde{\phi}_L$. Right: adjoint stream function $\tilde{\psi}_L$.

### 7.2 Airfoil with Finite Trailing-Edge Angle

The second case involves incompressible potential flow at zero incidence ($\alpha$ = 0º) around a symmetric van de Vooren airfoil given by the conformal transformation [30]

$$z = \frac{(\zeta - R)^k}{(\zeta - \sigma R)^{k-1}} + 1, \tag{74}$$

where $R = (1+\sigma)^{k-1} / 2^k$, $\sigma$ is a thickness parameter and $k$ is related to the trailing-edge angle $\tau$ as $k = 2 - \tau / \pi$. The transformation (74) maps the airfoil in the $z$ plane to a circle of radius $R$ centered at the origin in the $\zeta$ plane. In this paper, we set $\sigma = 0.0371$ and $k = 86/45$, resulting in an airfoil with 12% thickness and finite trailing edge angle $\tau = 16°$ shown in Figure 2, where contour maps of the analytic lift-based adjoint solutions are also presented. The analytic flow solution can be obtained with conformal transformation techniques [30]. The velocity in the plane of the airfoil is given in terms of the velocity in the circle plane (72) and the derivative of the conformal transformation (74) as

$$\begin{aligned} u_{airfoil} &= \mathrm{Re}(\partial\zeta / \partial z) u_{circle} + \mathrm{Im}(\partial\zeta / \partial z) v_{circle} \\ v_{airfoil} &= \mathrm{Re}(\partial\zeta / \partial z) v_{circle} - \mathrm{Im}(\partial\zeta / \partial z) u_{circle} \end{aligned} \tag{75}$$

The adjoint variables are obtained from (43), (34) and (75) as

$$\begin{aligned} \tilde{\phi}_L(x,y) &= \frac{\rho}{c_\infty} v_{airfoil}(x,y) - \frac{\rho}{c_\infty} q_\infty \frac{2R\,\mathrm{Im}\left(\zeta(x+iy)\right)}{\left|\zeta(x+iy) - R\right|^2} \\ \tilde{\psi}_L(x,y) &= -\frac{\rho}{c_\infty} u_{airfoil}(x,y) + \frac{\rho}{c_\infty} q_\infty \frac{\left|\zeta(x+iy)\right|^2 - R^2}{\left|\zeta(x+iy) - R\right|^2} \end{aligned} \tag{76}$$

As in the circle case, the analytic adjoint solutions show singularities at the trailing edge at $z$ = 1 (which corresponds to $\zeta(1) = R$ in the circle plane.)

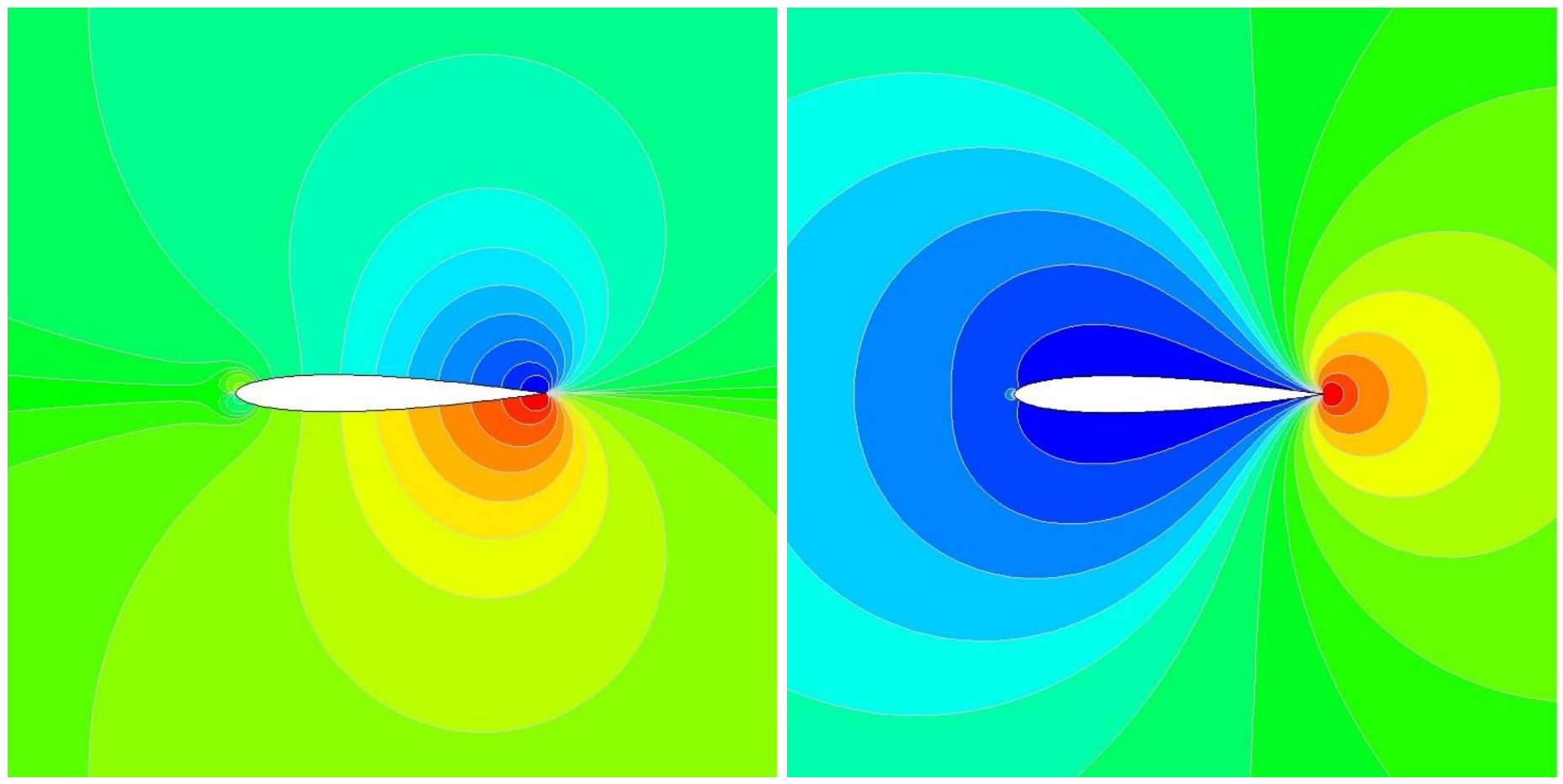

**Figure 2.** Analytic lift-based adjoint solution for incompressible potential flow with zero incidence ($\alpha$ = 0º) around a van de Vooren airfoil with trailing-edge angle $\tau = 16°$ and 12% thickness. Left: adjoint potential $\tilde{\phi}_L$. Right: adjoint stream function $\tilde{\psi}_L$.

## 8 Conclusions

In this paper we have revisited the adjoint equations for two-dimensional potential flow, for which a complete analytic formulation including lifting flows is still missing. As a step towards a continuous adjoint formulation, we have obtained and analyzed the analytic adjoint potential solutions for incompressible flow when the cost function is the aerodynamic force projected along a certain direction. Since potential flows can be described by either a potential or a stream function, the corresponding adjoint problems can be formulated in terms of an adjoint potential or stream function as well, which in the incompressible case turn out to be the imaginary and real parts, respectively, of an analytic function. The analytic adjoint solutions can be obtained by the Green's function approach, unveiling that the analytic solution for the adjoint potential is the linearized cost function corresponding to a point mass source, while the analytic solution for the adjoint stream function is the linearized cost function corresponding to a point vortex. This conclusion is confirmed by matching the adjoint potential and stream functions with the adjoint variables of the incompressible Euler equations, which yields another way to obtain the analytic solutions.

For aerodynamic lift, the analytic adjoint solutions contain two functions whose origin is in the perturbed Kutta condition. These functions, which turn out to be the Poisson kernel and its harmonic conjugate, are singular at the trailing edge or rear stagnation point, giving rise to the well-known adjoint singularity. Using the properties of these functions and their relation to the Kutta condition, we make the observation that they can only be obtained if the adjoint wall boundary conditions contain singular forcing terms involving the Dirac delta function. We further show that these terms can be explained, without the need to address the cutting line, which in any case is missing from the stream function formulation, if the linearized lift contains an additional term involving the perturbation velocity at the t.e. This term enforces the Kutta condition by replacing the unknown circulation appearing in the linearized force integral by its correct value.

There are several possible avenues for clarifying and extending the results of this paper. One involves finding a boundary formulation independent of conformal mappings, along the lines of the analysis of the trailing edge condition in [26], for example. Another possible avenue should aim at better understanding the formulation of the Kutta condition put forward in the paper. Finally, it would be interesting to extend these results to compressible potential flow. We plan to address some of these issues in future work.

## Funding

The research described in this paper has been supported by INTA and the Ministry of Defence of Spain under the grant IDATEC (IGB21001).

## Data Availability

The data that support the findings of this study are available from the corresponding authors upon reasonable request.